\documentstyle[twoside,fleqn,espcrc2]{article}
\def \MSbar {\vbox{\hrule\kern 1pt\hbox{\rm MS}}}
\font \smallrm = cmr7 at 8pt
\def \sMSbar {\vbox{\hrule\kern 1pt\hbox{\smallrm MS}}}
\def \GeV { {\ \rm GeV} }
\def\red#1{#1}
\def\blue#1{#1}
\def\green#1{#1}
\def\sienna#1{#1}
\def\magenta#1{#1}

\input epsf.tex
\def\DESepsf(#1 width #2){\epsfxsize=#2 \epsfbox{#1}}
\title{Parton Distribution Functions}

\author{Davison E.\ Soper
\address{Institute of Theoretical Science, 
         University of Oregon, \\ 
         Eugene, Oregon 97403, USA}%
\thanks{Research supported by U.\ S.\  Department of Energy
        grant DE-FG03-96ER40969.}
}
       
\begin{document}

\begin{abstract}
\noindent
Talk at Lattice 96 Conference, St.\ Louis, June 1996.

\bigskip
Parton distribution functions give the probability to find partons
(quarks and gluons) in a hadron as a function of the fraction $x$ of
the proton's momentum carried by the parton.  They are conventionally
defined in terms of matrix elements of certain operators. They are
determined from experimental results on short distance scattering of
the partons. Integrals of these functions weighted with $x^n$ are
calculable using lattice QCD. Some simple models for their behavior
have implications that could also be tested in lattice QCD.
\end{abstract}

\maketitle

\section{Intuitive meaning of parton distribution functions}

\begin{figure}[htb]
\vspace{9pt}
\centerline{\DESepsf(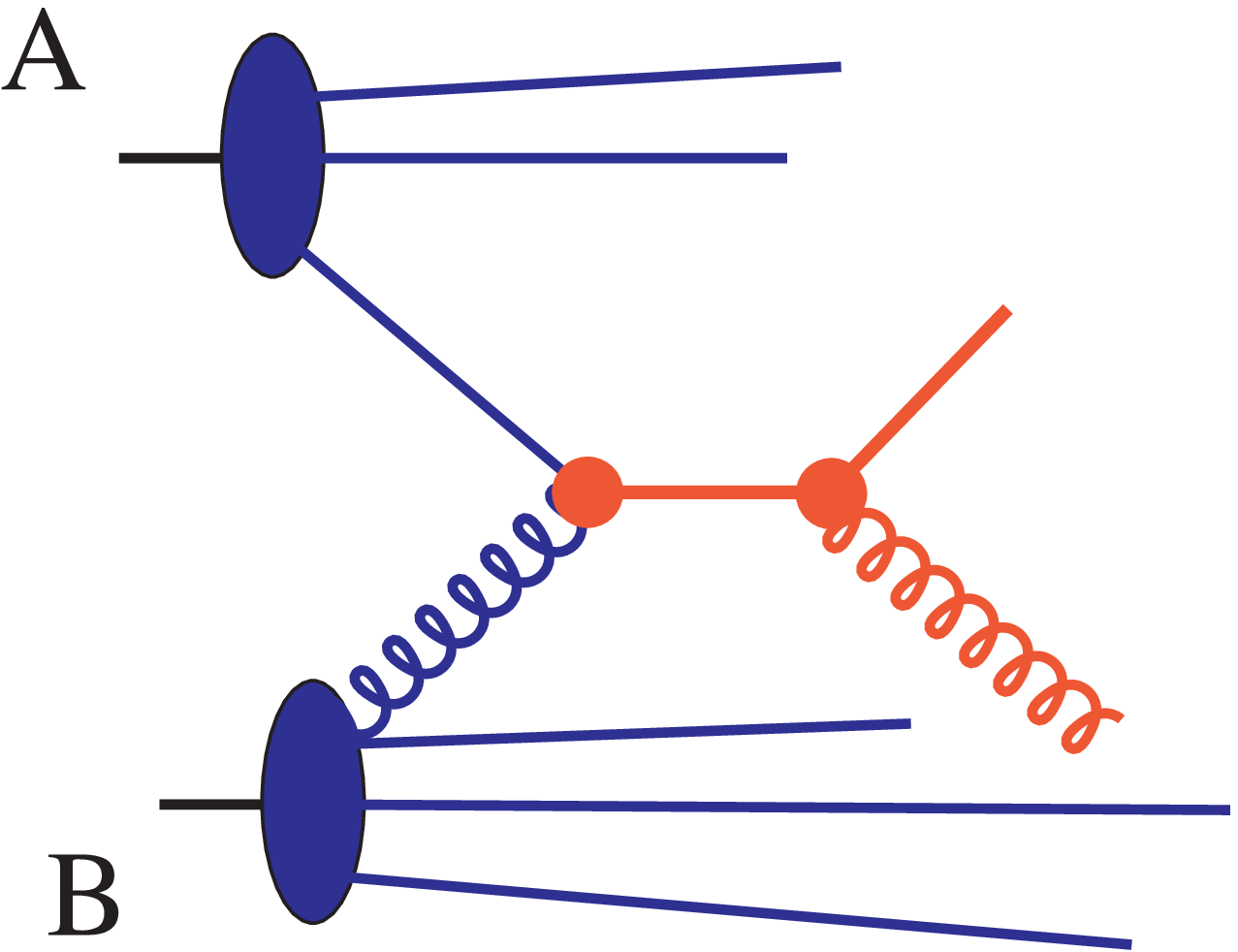 width 5 cm)}
\caption{Hadron $A$ + hadron $B$ $\to $ 2 partons.}
\label{fig:jetgraph}
\end{figure}

Let $d\sigma$ be a cross section involving short distances. For
instance, we may consider the process hadron $A$ + hadron $B$ $\to $
jet + $X$ at the Fermilab collider. Let the jet have a high transverse
momentum $P_T$. Intuitively, the observed jet begins as a single quark
or gluon that emerges from a parton-parton scattering event with large
$P_T$, as illustrated in Fig.~\ref{fig:jetgraph}. (Typically, this
parton recoils against a single parton that carries the opposite
$P_T$.) The large $P_T$ parton fragments into the observed jet of
hadrons.

The physical picture illustrated in Fig.~\ref{fig:jetgraph} suggests
how we may write the cross section to produce the jet as a product of
three factors. A parton of type $a$ comes from from a hadron of type
$A$. It carries a fraction $x_A$ of the hadron's momentum. The
probability to find it is given by $\blue{f_{a/A}(x_A)}\, dx_A$. A
second parton of type $b$ comes from a hadron of type B. It carries a
fraction $x_B$ of the hadron's momentum. The probability to find it is
$\blue{f_{b/B}(x_B)}\, dx_B$. The functions $f_{a/A}(x)$ are the
parton distribution functions that are the subject of this talk.
The third factor is the cross section for the partons to make the
observed jet, $\red{d\hat\sigma}$. This parton level cross section
is calculated using perturbative QCD.


\subsection{Factorization}

\begin{figure}[htb]
\vspace{9pt}
\centerline{\DESepsf(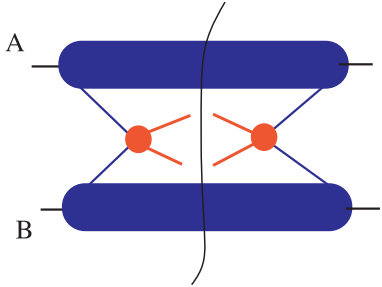 width 5 cm)}
\caption{Factorization for hadron collisions.}
\label{fig:factorization}
\end{figure}

We have been led by the intuitive parton picture of
Fig.~\ref{fig:jetgraph} to write the cross section for jet production
in the following form
\begin{eqnarray}
\lefteqn{{d\sigma \over d P_T} \sim}
\label{factor}\\
&&{\hskip - 0.7 cm}
\sum_{a,b}\int\! dx_A\,
\blue{f_{a/A}(x_A,\mu)}\
\int\! dx_B\, \blue{f_{b/B}(x_B,\mu)}\
\red{{d\hat\sigma \over d P_T}}.
\nonumber
\end{eqnarray}
Here the parton level cross section has a well behaved expansion
in powers of $\alpha_s$,
\begin{equation}
\red{{d\hat\sigma \over d P_T}} \sim 
\sum_N \left(\!\alpha_s(\mu)\over\pi\right)^{\!N}\!
H_N(x_A,x_B,P_T;a,b;\mu).
\end{equation}
The coefficients $H_N$ are calculable in perturbative QCD.

The principle of {\it factorization} asserts that
Eq.~(\ref{factor}) holds up to corrections of order

\smallskip
$\bullet$  
$(m/P_T)^n$  where $m$ is a typical hadronic mass

\hskip 0.4 cm scale and the power $n$ depends on the 

\hskip 0.4 cm process, and

$\bullet$
$\left(\alpha_s(\mu)\right)^{\!L}$ from truncating 
the expansion of 

\hskip 0.4 cm $d\hat\sigma/d P_T$.

\smallskip
\noindent
For our purposes, we can regard factorization as an established
theorem of QCD, although this subject is not without its loose ends.
A review may be found in Ref.~\cite{theorems}.

As we have seen, Eq.~(\ref{factor}) has a simple intuitive meaning.
However, the appearance of a parameter $\mu$ in Eq.~(\ref{factor})
hints that there is more to the equation than just a model. The
parameter $\mu$, which has dimensions of mass, is related to the
renormalization of the strong coupling $\alpha_s(\mu)$ and of the
operators in the definition of the parton distribution functions
$f_{a/A}(x_A,\mu)$. (Often, one uses two separate parameters in these
two places.)

At the Born level, the parton level cross section $\red{d\hat\sigma/d
P_T}$ is calculated in a straightforward manner. At the
next-to-leading order and beyond, the calculation is not so
straightforward. Various divergences appear in a naive calculation.
The divergences are removed and the dependence on the scale $\mu$
appears in their place. The precise rules for calculating
$\red{d\hat\sigma/d P_T}$ follow once you have set the definition of
the parton distribution functions $\blue{f_{a/A}(x,\mu)}$. These rules
enable one to do practical calculations. For example, for jet
production, the first two terms in $\red{d\hat\sigma/d P_T}$ are known
in the form of computer code that puts together the pieces of
Eq.~(\ref{factor}) and produces a numerical cross section
\cite{jetxsect}.


\subsection{Reality Check}

\begin{figure}[htb]
\vspace{9pt}
\centerline{\DESepsf(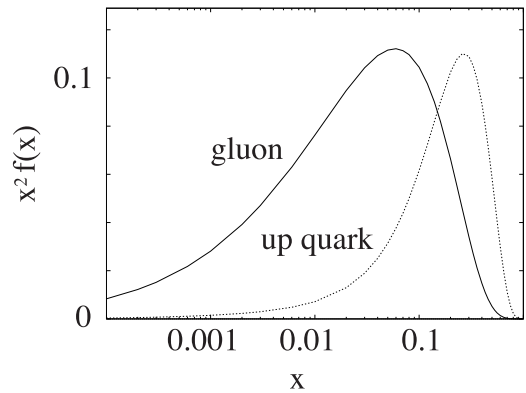 width 7cm)}
\caption{Gluon and up quark distributions in the proton according to
the CTEQ3M parton distribution set.}
\label{fig:cteq3m}
\end{figure}

Sets of parton distribution functions, one function for each
kind of parton in a proton, are produced to fit experiments. We will
examine the fitting process in a later section.
Fig.~\ref{fig:cteq3m} is a graph of the gluon distribution and the
up-quark distribution in a proton, according to a parton
distribution set designated CTEQ3M \cite{cteq3m}. The figure
\cite{potpourri} shows $x^2 f_{a/A}(x,\mu)$ for $a = g$ and $a = u$,
$A = p$. Note that
\begin{equation}
\int_0^1 dx\ x\, f_{a/A}(x,\mu) = 
\int d\log x\ x^2f_{a/A}(x,\mu),
\end{equation}
so the area under the curve is the momentum fraction carried by
partons of species $a$.

\subsection{Significance}

\begin{figure}[htb]
\vspace{9pt}
\centerline{\DESepsf(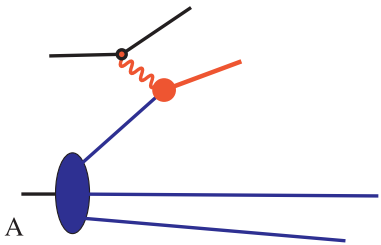 width 5cm)}
\caption{Deeply inelastic scattering.}
\label{fig:dis}
\end{figure}

Knowledge of parton distribution functions is necessary for the
description of hard processes with one or two hadrons in the initial
state. With two hadrons in the initial state, as at Fermilab or the
future Large Hadron Collider, observed short distance cross sections
take the form
\begin{eqnarray}
\lefteqn{{d\sigma} \sim}
\label{onehadron}\\
&&{\hskip - 0.4 cm}\sum_{a,b}\int\! dx_A\, \blue{f_{a/A}(x_A,\mu)}\
\int\! dx_B\, \blue{f_{b/B}(x_B,\mu)}\
\red{{d\hat\sigma}}.
\nonumber
\end{eqnarray}
With one hadron in initial state, as in deeply inelastic lepton
scattering at HERA (Fig.~\ref{fig:dis}), the cross section has
the form
\begin{equation}
{d\sigma} \sim
\sum_{a}\int dx_A\, \blue{f_{a/A}(x_A,\mu)}\
\red{{d\hat\sigma}}.
\label{twohadrons}
\end{equation}
In either case, one has no predictions without knowledge of the
parton distribution functions.

The essence of Eqs.~(\ref{onehadron}) and (\ref{twohadrons}) above is
that in high energy, short distance collisions a hard scattering
probes the system quickly, while the strong binding forces act slowly.
Thus one needs to know probabilities to find partons in a fast moving
hadron as seen by an approximately instantaneous probe. This is the
information encoded in the parton distribution functions. It should be
evident that this information is not only useful, but also valuable 
in understanding hadron structure. Parton distribution functions are
not everything, however. They provide a relativistic view only, a
view quite different from the view that might be most economical for
the description of a hadron at rest. Furthermore, they provide no
information on correlations among the partons.


\section{Translation to operators}

We are now ready to examine the technical definition of parton
distribution functions. There are, in fact, two definitions in current
use. I will describe the \MSbar\ definition, which is the most
commonly used. There is also a DIS definition, in which deeply
inelastic scattering plays a privileged role. The interested reader
may consult the CTEQ Collaboration's {\it Handbook of Perturbative
QCD} \cite{handbook} for information on the DIS definition. There are
also different ways to think about the \MSbar\ definition. In this
talk, I define the parton distributions directly in terms of field
operators along a light-like line, as in \cite{partondef}. An
equivalent construction from a different point of view may be found
in \cite{altpartondef}. As we will see, moments of the parton
distribution functions are related to matrix elements of certain
local operators, which appear in the operator product expansion for
deeply inelastic scattering. This relation could be also used as the
definition.

I distinguish between the {\it parton distribution functions}
$f_{a/A}(x,\mu)$ and the {\it structure functions} $F_1(x,Q^2)$,
$F_2(x,Q^2)$, and $F_3(x,Q^2)$ that are measured in deeply inelastic
lepton scattering.

\subsection{Null coordinates}

I will use null plane coordinates and momenta defined by
\begin{equation}
x^\pm = (x^0 \pm x^3)/\sqrt 2\,,
\hskip 0.3 cm
P^\pm = (P^0 \pm P^3)/\sqrt 2\,.
\end{equation}

\begin{figure}[htb]
\vspace{9pt}
\centerline{\DESepsf(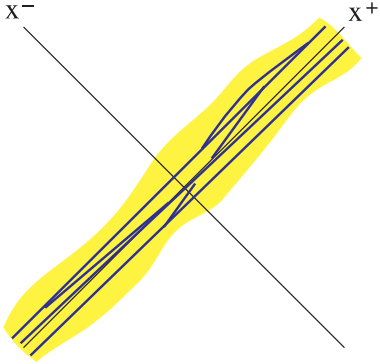 width 5cm)}
\caption{World lines of the partons in a fast-moving proton.}
\label{fig:lightcone}
\end{figure}

Imagine a proton with a big $P^+$, a small $P^-$, and $\vec  P_T = 0$.
The partons in such a proton move roughly parallel to the $x^+$ axis,
as illustrated in Fig.~\ref{fig:lightcone}. One can treat $x^+$ as
``time,'' so that the system propagates from one plane of equal $x^+$
to another. For our fast moving proton, the interval in $x^+$ between
successive interactions among the partons is typically large.

Notice that the invariant dot product between $P^\mu$ and $x^\mu$ is
\begin{equation}
P\cdot x = P^+ x^- + P^- x^+ - \vec P_T\cdot \vec x_T\,.
\end{equation}
Thus the generator of ``time'' translations is $P^-$.


\subsection{Null plane field theory}

We will define parton distribution functions by taking a snapshot of
the proton on a plane of equal $x^+$ in Fig.~\ref{fig:lightcone}.
To motivate the definition, we use field theory quantized on planes of
equal $x^+$ \cite{KS}. This quantization uses the gauge $A^+ = 0$.
Then the unrenormalized quark field operator $\psi_0$ is expanded in
terms of 

\smallskip
$\bullet$ quark destruction operators $b$ and

$\bullet$ antiquark creation operators $d^\dagger$

\smallskip\noindent
using simple spinors $w(s)$ normalized to $w^\dagger w = 1$:

\begin{eqnarray}
\lefteqn{{\scriptstyle{1\over 2}}\gamma^-\gamma^+
\sienna{\psi_0}(x^+,x^-,\vec x_T) =} 
\\
&&\hskip - 0.3 cm
{ 1 \over (2\pi)^3}\int_0^\infty {d k^+ \over 2 k^+}\, 
\int d \vec k_T
\sum_s (\sqrt 2 k^+)^{1/2}
\nonumber\\
&&\hskip - 0.3 cm \times\biggl\{
e^{-i\,(k^+ x^- - \vec k_T\cdot \vec x_T)}\,
w(s)\, \green{b}(k^+,\vec k_T;s;x^+)
\nonumber\\
&& {}+
e^{+i\,(k^+ x^- - \vec k_T\cdot \vec x_T)}\,
w(-s)\, \green{d^\dagger}(k^+,\vec k_T;s;x^+)
\biggr\}.
\nonumber
\end{eqnarray}
The factor ${\scriptstyle{1\over 2}}\gamma^-\gamma^+$ here serves to
project out the components of the quark field $\psi_0$ that are the
independent dynamical operators in null plane field theory.


\subsection{The quark distribution function}

We are now ready to define the (unrenormalized) quark distribution
function. Let $|P\rangle$ be the state vector for a hadron of type $A$
carrying momentum $P^\mu$. Take the hadron to be spinless in order
to simplify the notation. Construct the unrenormalized distribution
function for finding quarks of flavor $j$ in hadron $A$ as
\begin{eqnarray}
\lefteqn{\blue{f^{(0)}_{j/A}(x)} \times
\langle P^{+\prime},\vec P_T^\prime | P^{+},\vec P_T \rangle =}
\nonumber\\
&&\hskip  0.5 cm{ 1 \over 2 x (2\pi)^3}\int d\vec k_T \sum_s\,
\langle P^{+\prime},\vec P_T^\prime |
\nonumber\\
&&\hskip  0.5 cm
\times
\green{b^\dagger_j}(xP^+,\vec k_T;s;x^+)\,
\green{b_j}(xP^+,\vec k_T;s;x^+)
\nonumber\\
&&\hskip  0.5 cm
\times
| P^{+},\vec P_T \rangle\ .
\label{quark0def}
\end{eqnarray}
In Eq.~(\ref{quark0def}) there are factors relating to the
normalization of the states and  the creation/destruction operators.
There is a quark number operator $\green{b^\dagger b}$ for flavor $j$.
We integrate over the quark transverse momentum $\vec k_T$ and sum
over the quark spin $s$.


\subsection{Translation to coordinate space}

With a little algebra, we find
\begin{eqnarray}
\lefteqn{\blue{f^{(0)}_{j/A}(x)} 
={ 1 \over 4\pi}\int dy^-
e^{-i xP^+y^-}}
\\
&&\hskip - 0.6cm \times\langle P^{+}\!,\vec 0_T |
\overline{\sienna{\psi}}_{0,j}(0,y^-,\vec 0_T)
\gamma^+
{\sienna{\psi}}_{0,j}(0,0,\vec 0_T)
| P^{+}\!,\vec 0_T \rangle.
\nonumber
\end{eqnarray}
Notice that the (still unrenormalized) quark distribution function is
an expectation value in the hadron state of a certain operator. The
operator is not local but ``bilocal.'' The two points, $(0,y^-,\vec
0_T)$ and $(0,0,\vec 0_T)$, at which the field operators are evaluated
are light-like separated. The formula directs us to integrate over
$y^-$ with the right factor so that we annihilate a quark with plus
momentum $xP^+$.


\subsection{Gauge invariance}

Before turning to the renormalization of the operator that occurs in
the quark distribution function, we note that the definition as it
stands relies on the gluon potential $A^\mu(x)$ being in the
gauge $A^+ = 0$. Let us modify the formula so that

\smallskip
$\bullet$ the operator is gauge invariant and

$\bullet$ we match the previous definition in 

\hskip 0.3 cm $A^+=0$ gauge.

\smallskip\noindent
The gauge invariant definition is
\begin{eqnarray}
\lefteqn{\blue{f^{(0)}_{j/A}(x)} 
={ 1 \over 4\pi}\int\! dy^-
e^{-i xP^+y^-}\ 
\langle P^{+}\!,\vec 0_T |}
\nonumber\\
&&\hskip 1.0cm
\times
\overline{\sienna{\psi}}_{0,j}(0,y^-,\vec 0_T)
\gamma^+ \magenta{{\cal O}_0}\
{\sienna{\psi}}_{0,j}(0,0,\vec 0_T)
\nonumber\\
&&\hskip 1.0cm \times
| P^{+}\!,\vec 0_T \rangle\ ,
\label{quarkdef2}
\end{eqnarray}
where
\begin{equation}
\magenta{{\cal O}_0} ={\cal P}
\exp\left(
ig_0 \int_0^{y^-}\!\!\! dz^-\, {\sienna{A}}_{0,a}^+(0,z^-,\vec 0_T)\,
t_a
\right).
\label{eikonal}
\end{equation}
Here ${\cal P}$ denotes a path-ordered product, while the $t_a$ are
the generators for the {\bf 3} representation of SU(3). There is an
implied sum over the color index $a$.


\subsection{Interpretation of the eikonal gauge operator}

The appearance of the operator $\cal O$, Eq.~(\ref{eikonal}), in 
the definition (\ref{quarkdef2}) seems to be just a technicality.
However, this operator has a physical interpretation that is of some
importance. Let us write this operator in the form
\begin{eqnarray}
\lefteqn{
\magenta{{\cal O}_0} = \overline{\cal P}
\exp\left(
-ig_0 \int_{y^-}^\infty\!\!\! dz^-\, {\sienna{A}}_{0,a}^+(0,z^-,\vec
0_T)\, t_a
\right)}
\nonumber\\
&&\times 
{\cal P}
\exp\left(
ig_0 \int_0^\infty\!\!\! dz^-\, {\sienna{A}}_{0,a}^+(0,z^-,\vec 0_T)\,
t_a
\right).
\label{eikonalmod}
\end{eqnarray}
Inserting this form in the definition (\ref{quarkdef2}), we can
introduce a sum over states $|N\rangle\langle N|$ between the
two exponentials in Eq.~(\ref{eikonalmod}). We take these states to
represent the final states after the quark has been ``measured.''

\begin{figure}[htb]
\vspace{9pt}
\centerline{\DESepsf(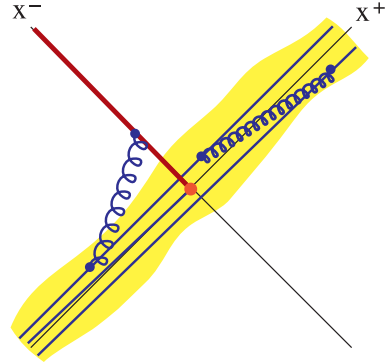 width 5cm)}
\caption{Effect of the eikonal gauge operator.}
\label{fig:lightcone3}
\end{figure}

Consider now a deeply inelastic scattering experiment that is used to
determine the quark distribution. The experiment doesn't just
annihilate the quark's color. In a suitable coordinate system, a quark
moving in the plus direction is struck and exits to infinity with
almost the speed of light in the minus direction, as illustrated in
Fig.~\ref{fig:lightcone3}. As it goes, the struck quark interacts with
the gluon field of the hadron.

We can now see that the role of the operator $\cal O$ is to replace
the struck quark with a fixed color charge that moves along a
light-like line in the minus-direction, mimicking the motion of the
actual struck quark in a real experiment.


\subsection{Renormalization}

We now discuss the renormalization of the operator products in
the definition (\ref{quarkdef2}). We use \MSbar\ renormalized fields
${\sienna{\psi}}(x)$ and ${\sienna{A}}^\mu(x)$ and we use the \MSbar\
renormalized coupling $g$. The field operators are evaluated at
points separated by $\Delta x$ with $ \Delta x^\mu \Delta x_\mu = 0$.
For this reason, there will be ultraviolet divergences from the
operator products. We elect to renormalize the operator products with
the \MSbar\ scheme.

\begin{figure}[htb]
\vspace{9pt}
\centerline{\DESepsf(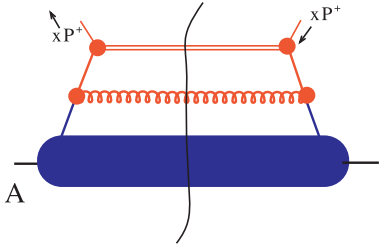 width 5cm)}
\caption{Renormalization of an ultraviolet divergent
loop integration.}
\label{fig:renormalize}
\end{figure}

For instance, Fig.~\ref{fig:renormalize} illustrates one of the
diagrams for the distribution of quarks in a proton. Before it is
measured, the quark emits a gluon into the final state. There is a
loop integration over the minus and transverse components of the
measured quark's momentum. This loop integration is ultraviolet
divergent. To apply \MSbar\ renormalization, we perform the
integration in $4-2\epsilon$ dimensions, including a factor $(\mu^2
e^{\gamma}/4\pi)^\epsilon$ that keeps the dimension constant while
supplying some conventional factors. The integral will consist of a
pole term proportional to $1/\epsilon$ plus terms that are finite as
$\epsilon \to 0$. We simply subtract the pole term. Notice that
\MSbar\ renormalization introduces a scale $\mu$.

The definition of the renormalized quark distribution function is thus
\begin{eqnarray}
\lefteqn{\blue{f_{j/A}(x,\mu)} 
={ 1 \over 4\pi}\int dy^-
e^{-i xP^+y^-}\
\langle P^{+},\vec 0_T |}
\nonumber\\
&&\hskip 1.0cm \times
\overline{\sienna{\psi}}_{j}(0,y^-,\vec 0_T)
\gamma^+ {\magenta{\cal O}}\
{\sienna{\psi}}_{j}(0,0,\vec 0_T)
\nonumber\\
&&\hskip  1.0cm \times
| P^{+},\vec 0_T \rangle _{{\red{\sMSbar}}}\,,
\label{quarkdef}
\end{eqnarray}
where the \MSbar\ denotes the renormalization prescription and
where
\begin{equation}
\magenta{{\cal O}}={\cal P}
\exp\left(
ig \int_0^{y^-}\!\!\! dz^-\, {\sienna{A}}_{a}^+(0,z^-,\vec 0_T)\, t_a
\right).
\label{eikonaldef}
\end{equation}
%


\subsection{Antiquarks and gluons}

We now have a definition of parton distribution functions for quarks.
For antiquarks, we use charge conjugation to define
\begin{eqnarray}
\lefteqn{
{\blue{f_{\bar j/A}(x,\mu)}} 
={ 1 \over 4\pi}\int dy^-
e^{-i xP^+y^-}\
\langle P^{+}\!\!,\vec 0_T |
}
\nonumber\\
&&\hskip 0.6 cm\times 
{\rm Tr}\!\left\{\!\gamma^+ {\sienna{\psi}}_{j}(0,y^-,\vec 0_T)
{\magenta{{\cal O}}}\
\overline{\sienna{\psi}}_{j}(0,0,\vec 0_T)\right\}\!
\nonumber\\
&&\hskip 0.6 cm\times
| P^{+}\!\!,\vec 0_T \rangle _{\sMSbar}\ ,
\label{antiquarkdef}
\end{eqnarray}
where
\begin{equation}
{\magenta{{\cal O}}}={\cal P}
\exp\left(
-ig \int_0^{y^-}\!\!\! dz^-\, {\sienna{A}}_{a}^+(0,z^-,\vec 0_T)\, t^T_a
\right).
\end{equation}

For gluons we begin with the number operator in $A^+ = 0$ gauge.
Proceeding analogously to the quark case, we obtain an expression
involving the field strength tensor $F_a^{\mu\nu}$ with color index
$a$: 
\begin{eqnarray}
\lefteqn{{\blue{f_{g/A}(x,\mu)}} 
={ 1 \over 2\pi\,xP^+}\int dy^-
e^{-i xP^+y^-}\
\langle P^{+}\!\!,\vec 0_T |}
\nonumber\\
&&\hskip 1.2cm \times
 {\sienna{F}}_{\!a}(0,y^-,\vec 0_T)
^{+\nu}{\magenta{{\cal O}}}_{ab}\
{\sienna{F}}_{\!b}(0,0,\vec 0_T)_\nu^{\ +}
\nonumber\\
&&\hskip 1.2cm \times
| P^{+}\!\!,\vec 0_T \rangle _{\sMSbar}\ ,
\label{gluondef}
\end{eqnarray}
where
\begin{equation}
{\magenta{{\cal O}}}={\cal P}
\exp\left(
ig \int_0^{y^-}\!\!\! dz^-\, {\sienna{A}}_{c}^+(0,z^-,\vec 0_T)\, t_c
\right).
\end{equation}
Here the $t_c$ generate the {\bf 8} representation of SU(3).


\section{Renormalization group}

A change in the scale $\mu$ induces a change in the parton
distribution functions $f_{a/A}(x,\mu)$. The change comes from the
change in the amount of ultraviolet divergence that renormalization
is removing. Since the operators are non-local in $y^-$, the
ultraviolet counterterms are integral operators in $k^+$ or
equivalently in momentum fraction $x$. Since the ultraviolet
divergences mix quarks and gluons, so do the counterterms.

One finds
\begin{eqnarray}
\lefteqn{\mu^2{ d \over d \mu^2}{\blue{f_{a/A}(x,\mu)}} =}
\nonumber\\
&&
\int_x^1 {d\xi\over \xi}\sum_b\ {\red{P_{a/b}(x/\xi,\alpha_s(\mu))}}\
{\blue{f_{b/A}(\xi,\mu)}}.
\label{APeqn}
\end{eqnarray}
The Altarelli-Parisi (= GLAP = DGLAP) kernel $P_{a/b}$ is expanded in
powers of $\alpha_s$. The $\alpha_s^1$ and $\alpha_s^2$ terms are
known and used.


\subsection{Renormalization group interpretation}

\begin{figure}[htb]
\vspace{9pt}
\centerline{\DESepsf(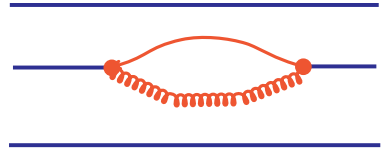 width 5cm)}
\caption{A quark can fluctuate into a quark plus a gluon in a
small space-time volume.}
\label{fig:scales}
\end{figure}

The derivation of the renormalization group equation (\ref{APeqn}) is
rather technical. One should not lose sight of its intuitive meaning.
Parton splitting is always going on as illustrated in
Fig.~\ref{fig:scales}. A probe with low resolving power doesn't see
this splitting. The renormalization parameter $\mu$ corresponds to the
physical resolving power of the probe. At higher $\mu$, field
operators representing an idealized experiment can resolve the
mother parton into its daughters.

\subsection{Renormalization group result}

\begin{figure}[htb]
\vspace{9pt}
\leftline{\DESepsf(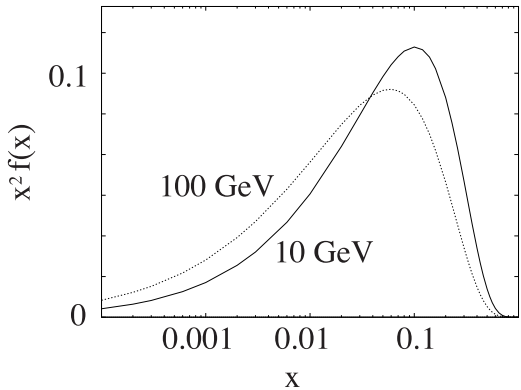 width 7cm)}
\caption{Evolution of the gluon distribution between 
$\mu = 10 \GeV$ and $\mu = 100 \GeV$.}
\label{fig:evolve}
\end{figure}

One can use the renormalization group equation (\ref{APeqn}) to 
find the parton distributions at a scale $\mu$ if they are known at a
lower scale $\mu_0$. Fig.~\ref{fig:evolve} shows an example, the gluon
distribution at $\mu = 10 \GeV$ and at $\mu = 100 \GeV$ (using the
CTEQ3M parton distribution set).  Notice that with greater resolution,
a gluon typically carries a smaller momentum fraction $x$ because of
splitting.

\section{Translation to local operators}

We have defined the parton distributions as hadron matrix elements of
certain operator products, where the operators are evaluated along a
light-like line. Now we relate the parton distributions to products of
operators all at the same point. It is these local operator products
that were originally used in the interpretation of deeply inelastic
scattering experiments. For lattice QCD, evaluation of operator
products at light-like separations would be, at best, very difficult,
so the translation to local operator products seems essential. 

\subsection{Quarks}

Note that, according to the definitions (\ref{quarkdef}) and
(\ref{antiquarkdef}), 
\begin{eqnarray}
f_{j/A}(x,\mu) &=& 0\,, \hskip 2 cm {\magenta{x>1}}\,,
\nonumber\\
f_{\bar j/A}(x,\mu) &=& 0\,, \hskip 2 cm {\magenta{x>1}}\,,
\nonumber\\
f_{j/A}({\blue{-x}},\mu) &=& {\blue{-}} f_{{\blue{\bar j}}/A}(x,\mu).
\end{eqnarray}
Consider the moments of the quark/antiquark distributions defined by
\begin{eqnarray}
\lefteqn{M_j^{(J)}(\mu) =}
\\
&& \int_0^1 { dx \over x}\, {\red{x^J}}
\left\{
f_{j/A}(x;\mu)
+{\blue{(-1)^J}} f_{{\blue{\bar j}}/A}(x;\mu)
\right\}
\nonumber
\end{eqnarray}
for $J = 1,2,\dots$. Given the properties above, this is
\begin{equation}
M_j^{(J)}(\mu) = \int_{{\magenta{-\infty}}}^{{\magenta{\infty}}}
{ dx \over x}\, {\red{x^J}}
f_{j/A}(x;\mu)\,.
\end{equation}
Obtaining $\int_{-\infty}^{\infty}dx$ is the essential step.

From the operator definitions of the $f$s, this is
\begin{eqnarray}
\lefteqn{
M_j^{(J)}(\mu) 
=}
\nonumber\\&& 
{ 1 \over 4\pi}\int dy^- \int_{-\infty}^{\infty}dx\ 
e^{-i xP^+{\red{y^-}}}\ 
\left(
{ -i \over P^+}
{\red{{ \partial \over \partial y^-}}}
\right)^{J-1}
\nonumber\\
&&\ \ \times\langle P|
\overline\psi_{j}({\red{y^-}})
\gamma^+ {\cal O}({\red{y^-}},0)\
\psi_{j}(0)
| P \rangle _{\sMSbar}\ .
\end{eqnarray}
Performing the $x$-integration gives a $\delta(y^-)$.
Thus we get a local operator. The ${\red{\partial/\partial y^-}}$
differentiates either the quark field or the exponential of gluon
fields in $\cal O$, Eq.~(\ref{eikonaldef}).  We find
\begin{eqnarray}
\lefteqn{M_j^{(J)}(\mu) =}
\\
&&\hskip - 0.3cm
{ 1 \over 2}(P^+)^{-J}\,
\langle P|
\overline\psi_{j}(0)
\gamma^+ 
\left(
i{\red{D^+}}
\right)^{J-1}
\psi_{j}(0)
| P \rangle _{\sMSbar}\,,
\nonumber
\end{eqnarray}
where
\begin{equation}
{\red{D^\mu}} =  { \partial \over \partial y_\mu}
- i g\, A^\mu_a(y)\, t_a\,.
\end{equation}

We have now related the moments of the quark distribution to products
of operators evaluated at the same point.  However, this is not yet
ready for the lattice because it would not be easy to differentiate
the operators with respect to $y^-$. To obtain a more useful
expression, consider ${\blue{\langle {\cal O}_j^{(J)}\rangle}}$
defined by
\begin{eqnarray}
\lefteqn{\{P^{\mu_1}P^{\mu_1}\cdots P^{\mu_J}\}_{\rm TS}\
{\blue{\langle {\cal O}_j^{(J)}\rangle}}
=}
\\
&&\hskip -0.5cm
{1 \over 2}
\langle P|
\overline\psi_{j}(0)\,
\{\gamma^{\mu_1} 
iD^{\mu_2}
\cdots
iD^{\mu_J}
\}_{\rm TS}\,
\psi_{j}(0)
| P \rangle _{\sMSbar}\ .
\nonumber
\end{eqnarray}
where TS denotes taking the traceless symmetric part of the tensor
enclosed.

Then
\begin{eqnarray}
\lefteqn{{\blue{\langle {\cal O}_j^{(J)}\rangle}}
= M_j^{(J)}(\mu) =}
\\
&&\int_0^1 { dx \over x}\, {\red{x^J}}
\left\{
{\blue{f_{j/A}(x;\mu)}}
+(-1)^J {\blue{f_{\bar j/A}(x;\mu)}}
\right\}.
\nonumber
\end{eqnarray}
This is our final result.

We can now imagine the following program for a lattice calculation.
One could measure ${\blue{\langle {\cal O}_j^{(J)}\rangle}}$ on the
lattice for $J=1,2,\dots$. Of course, this would not be so easy, but
it would gives moments of the quark/antiquark distributions that
could be compared to the corresponding moments of the distributions
determined from experiments.


\subsection{Gluons}

We follow a similar analysis to relate the gluon distribution
function to local operator products. From the definition
(\ref{gluondef}), it follows that
\begin{eqnarray}
f_{g/A}(x,\mu) &=& 0\,, \hskip 2 cm {\magenta{x>1}}\,,
\nonumber\\
f_{g/A}({\blue{-x}},\mu) &=& {\blue{-}} f_{g/A}(x,\mu)\,.
\end{eqnarray}
For $J= 2,4,6,\dots$, we relate moment integrals over the interval
$0<x<1$ to moment integrals over all $x$
\begin{equation}
\int_0^1\! { dx \over x}\, {\red{x^J}}
f_{g/A}(x;\mu)
=
{1 \over 2}\int_{{\magenta{-\infty}}}^{{\magenta{\infty}}}\!
{ dx \over x}\, {\red{x^J}} f_{g/A}(x;\mu).
\end{equation}
Define operator matrix elements ${\blue{\langle {\cal
O}_g^{(J)}\rangle}}$ for $J = 2,4,6,\dots$ by
\begin{eqnarray}
\lefteqn{\{P^{\mu_1}P^{\mu_1}\cdots P^{\mu_J}\}_{\rm TS}\
\langle {\cal O}_g^{(J)}\rangle
=}
\\
&&\hskip -0.6cm
{1 \over 2}
\langle P|\{
F^{\mu_1\!\nu}\!(0)\,
iD^{\mu_2}
\cdots
iD^{\mu_{\!J-1}}\,
F_\nu^{\ \mu_{\!J}}\!(0)\,
\}_{\rm TS}\,
| P \rangle _{\sMSbar}\,.
\nonumber
\end{eqnarray}
Then these operator matrix elements are the moments of the gluon
distribution:
\begin{equation}
{\blue{\langle {\cal O}_g^{(J)}\rangle}}
= M_g^{(J)}(\mu)
=\int_0^1 { dx \over x}\, {\red{x^J}}
{\blue{f_{g/A}(x;\mu)}}\,.
\end{equation}
Again, one can contemplate measuring ${\blue{\langle {\cal
O}_g^{(J)}\rangle}}$ on the lattice for $J=2,4,\dots$. The result
could be compared to the corresponding moments of the
phenomenological gluon distribution.


\section{Determination from experiments}

I have alluded several times to the fact that parton distribution
functions are related to experimental results. Let us look at this in
some more detail.


\subsection{Example}

\begin{figure}[htb]
\centerline{\DESepsf(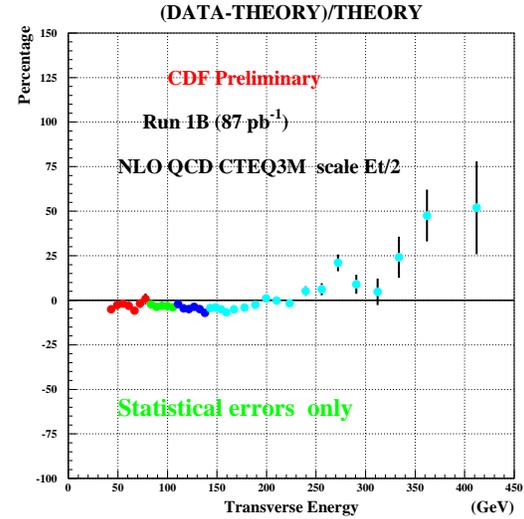 width 8 cm)}
\vskip - 2.5 cm
\caption{One jet inclusive jet cross section $d\sigma/dE_T$,
(Data $-$ Theory)/Theory versus $E_T$. Source: CDF.}
\label{fig:cdfjet}
\end{figure}

Fig.~\ref{fig:cdfjet} shows data for the cross section to observe one
jet plus anything in high energy $p\bar p$ collisions. The data from
the CDF Collaboration \cite{CDFjet} are compared to theory using
CTEQ3M parton distributions that did {\it not} have this data as
input. (In the figure, $E_T$ denotes essentially the transverse
momentum of the jet.)

\begin{figure}[htb]
\vspace{9pt}
\centerline{\DESepsf(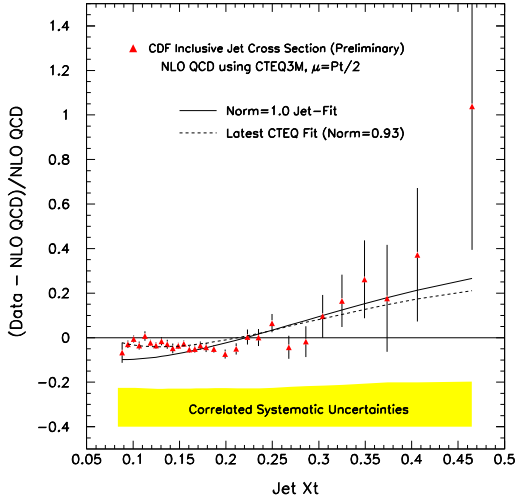 width 8 cm)}
\caption{(Data $-$ Theory)/Theory versus $x_T = 2 E_T/\sqrt s$
for CTEQ4HJ partons, ``Norm = 1.0 Jet-Fit.'' Source: CTEQ.}
\label{fig:newjet}
\end{figure}

One notes that the data do not agree with the theory at the highest
values of $E_T$. Does this indicate a breakdown of QCD at the smallest
distances? An alternative explanation \cite{CTEQjet} is that the gluon
distribution used in the theory was actually poorly constrained at
large $x$ ($x \sim 0.5$) by other experiments. In support of this
explanation, the CTEQ collaboration offered the theory curve shown in
Fig.~\ref{fig:newjet}. This theory curve was obtained using modified
parton distributions that still fit other data well. (Note: the data
in this figure predates that in Fig.~\ref{fig:cdfjet} and has lower
statistics.)

Could  a lattice measurement of moments of the gluon distribution
distinguish between these possibilities? In Table \ref{tab:moments},
I show the lowest moments of the gluon distributions from the two
parton distribution sets in question. The two gluon distributions are
very different at large $x$, but almost the same at small $x$. Thus
one has to go to the $J=6$ moment to see a substantial difference.
Measurement of $M^{(6)}_g(2 \GeV)$ in a lattice calculation would, I
think, be quite difficult.

\begin{table}[hbt]
\caption{Moments of gluon distribution for two sample parton
distributions sets}
\label{tab:moments}
\begin{tabular*}{7.5cm}{@{}@{\extracolsep{\fill}}rrr}
\hline
    \multicolumn{1}{r}{} 
  & \multicolumn{1}{r}{CTEQ3M} 
  & \multicolumn{1}{r}{CTEQ4HJ}  \\
\hline
$M^{(2)}_g(2 \GeV)$ &
$4.5 \times 10^{-1}$  &
$4.3 \times 10^{-1}$  \\ 
$M^{(4)}_g(2 \GeV)$   &
$1.6 \times 10^{-2}$  &
$1.9 \times 10^{-2}$  \\   
$M^{(6)}_g(2 \GeV)$   &
$2.1 \times 10^{-3}$  &
$4.4 \times 10^{-3}$  \\ 
\hline
\end{tabular*}
\end{table}


\subsection{What parton people do}

There are three main groups currently doing ``global fitting'' of
parton distributions:

$\bullet$ Martin, Roberts and Stirling (MRS) \cite{MRS},

$\bullet$ Gluck, Reya, and Vogt (GRV) \cite{GRV},

$\bullet$ Botts, Huston, Lai, Morfin, Owens, Qiu, 

\hskip 0.3 cm Tung,  Weerts,\dots (CTEQ) \cite{cteq3m}.

\noindent
Other groups have done this in the past, {\it e.g.} Duke and
Owens and Eichten, Hinchliffe, Lane and Quigg.

The idea of the global fitting is to adjust the parton distribution
functions to make theory and experiment agree for a wide range of
processes. For example, recent CTEQ fits have used the following
processes:

\smallskip

$e + p \to  X$ 

$\mu + p \to  X$ \hskip 2 cm $\mu + {}^2\!H \to  X$

$\nu + Fe \to  X$ \hskip 1.75 cm $\bar\nu + Fe \to  X$

$p + p \to \mu + \bar \mu + X$ \hskip 0.7 cm $p + {}^2\!H \to \mu +
\bar
\mu + X$

$p + Cu \to \mu + \bar \mu + X$

$p + \bar p \to W \to \ell + \bar \ell + X$

$p + \bar p \to \gamma + X$ \,.

\smallskip

The main features of a program of global fitting are as follows. One
chooses a starting scale $\mu_0$ (say 2 \GeV). Then one writes the
$f_{a/p}(x;\mu_0)$ in terms of several parameters for $a = g,u,\bar u,
d, \bar d, s, \bar s$. (Typically the heavy quark distributions, for $a
= c,\bar c, b, \bar b, t, \bar t$, are generated from evolution, not
fit to data.) For example, one may choose
\begin{equation}
f(x;\mu_0) = A x^B (1-x)^C (1 + E x^D)\,.
\end{equation}
The parton distribution functions obey certain flavor and momentum
sum rules, such as 
\begin{eqnarray}
\int_0^1 dx\ \left[
f_{u/p}(x,\mu) - f_{\bar u/p}(x,\mu)
\right] &=&2,
\nonumber\\
\sum_a\int_0^1 dx\  x
f_{a/p}(x,\mu)  
&=&1.
\end{eqnarray}
These sum rules follow from the operator definitions
(\ref{quarkdef}), (\ref{antiquarkdef}), (\ref{gluondef}). The
parametrizations are chosen so that the flavor and momentum sum rules
are obeyed exactly.

Now one picks a trial set of parameters. This determines
$f(x;\mu_0)$, from which one calculates $f(x;\mu)$ for all
$\mu$ by evolution. Next, given the $f(x;\mu)$, one generates theory
curves for each type of experiment used. Finally, one compares the
results to the data.

This sequence is iterated, adjusting the parameters to get a good fit.
There are more than 1000 data and only about 25 parameters to be fit,
so the fact that this procedure works at all is an indication that
QCD is the correct theory of the strong interactions.


\section{Relation to lattice QCD}

Let me give here a personal view about how lattice calculations of
the lowest moments of parton distributions, continuing the work
reported at this conference \cite{latticecalc}, might enhance our
understanding of particle physics. 

First, what do we know now? To a reasonable approximation, the parton
distribution functions have already been determined from experiment.
Furthermore, there is partial understanding of parton distributions in
two limits: one has some idea about $x \to 0$ behavior from Regge
physics and from the evolution equation;  one has some idea about
$x \to 1$ behavior from Feynman graphs. However, this understanding
is of a rather limited nature. In general there is little theoretical
understanding of why parton distributions are what they are.

A lattice calculation of moments of parton distributions could,
first of all, test QCD. One could calculate something about hadron
structure instead of simply measuring that structure. It is
possible, in principle, that lattice calculations could add to the
precision of parton determinations, particularly for the gluon
distribution. However it seems to me that the goal of doing better
by calculation than by experiment is rather distant at this point.
Finally, and perhaps most importantly, it seems to me that lattice
calculations might lead to an improved understanding of hadron
structure by either validating or falsifying simple pictures about
how the partons are arranged in a hadron. I turn to this question in
the sections that follow.


\section{Are partons inside ``valence quarks''?}

\begin{figure}[htb]
\vspace{9pt}
\centerline{\DESepsf(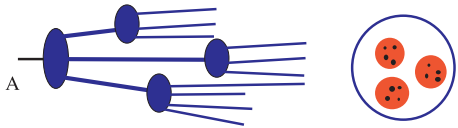 width 6 cm)}
\caption{A hadron may contain valence quarks that in turn
are made of short distance quarks and gluons.}
\label{fig:valon}
\end{figure}

Figure \ref{fig:valon} illustrates a simple picture of how partons
might be arranged in a proton. A proton consists of three valence
quarks, U U D. A valence quark in this sense is a big object. Each
valence quark consists of short distance quarks and gluons. I do not
know the history of this simple idea, but one illustrative paper on
the subject is Ref.~\cite{valon}.

This idea may be expressed in the formula
\begin{equation}
{\blue{f_{a/A}(x;\mu)}} = 
\int_x^1\! {d \xi \over \xi}\!\sum_{J = U,D}\! 
{\sienna{F_{J/A}(\xi)}}\,
{\green{f_{a/J}(x/\xi;\mu)}}.
\end{equation}
Here $F_{J/A}(\xi)$ represents the distribution of valence quarks of
flavor $J$ in a hadron of type $A$. Then $f_{a/J}(x/\xi;\mu)\,
d(x/\xi)$ is the probability to find a parton of type $a$ carrying a
fraction $x/\xi$ of the valence quark's momentum. 

Note that, without an additional hypothesis for
${\sienna{F_{J/A}(\xi)}}$, this formula has negative predictive
value: in order to predict the values of the functions
$f_{a/p}(x;\mu)$ we introduce unknown functions  $f_{a/J}(x/\xi;\mu)$
and $F_{J/A}(\xi)$. However, if you look not just at hadrons $p$ and
$n$ but at hadrons $p,n,\pi,\rho,\omega,\dots$ then there {\it is}
predictive power. There is experimental information on the
distribution of hadrons in the pion from $\pi + p \to \mu^+ + \mu^- +
X$. In addition, there is some experimental information on the
distribution of hadrons in the $\rho^0$ from deeply inelastic
scattering from photons. However, experimental information on the
distribution of partons in other hadrons is lacking. Here one sees an
advantage of calculation: it is much easier to create all kinds of
hadrons on a lattice than in an accelerator.


\subsection{Possibilities for $F_{J/A}(x)$}

One simple hypothesis for $F_{J/A}(x)$ would be that
a constituent quark is simply an elementary quark as seen at 
a very low resolution scale $\mu$. Thus one might identify
${\sienna{F_{J/A}(x)}} =  {\blue{f_{J/A}(x, 0.3 \GeV)}}$. However,
this hypothesis does not seem to work \cite{GRV}.

Another possibility is that the constituent quarks are very weakly
bound in a hadron. Then
\begin{eqnarray}
{\sienna{F_{U/p}(x)}}&\approx& 2\,\delta(x - 1/3),
\nonumber\\
{\sienna{F_{D/p}(x)}}&\approx& 1\,\delta(x - 1/3).
\end{eqnarray}
If one investigates only protons and neutrons, then this
hypothesis has zero predictive power. (That is better, of course,
than negative predictive power.) However, of one looks at
$p,n,\rho,\omega ...$ then there {\it is} some predictive power.  


\subsection{Parton correlations}

\begin{figure}[htb]
\vspace{9pt}
\centerline{\DESepsf(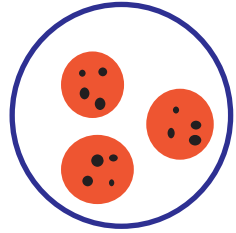 width 3 cm)}
\caption{If the short distance partons are contained
inside valence quarks, then the probability that two 
partons are close together is larger.}
\label{fig:valon2}
\end{figure}

The valence quark picture appears to say something about multiparton
correlations. Examining Fig.~(\ref{fig:valon2}), we see that the
partons in a proton are clumped together more than they would be if
they were spread throughout the proton. Thus a suitably defined two
parton correlation function would be larger at short distances than
it would be in an uncorrelated model.

\begin{figure}[htb]
\vspace{9pt}
\centerline{\DESepsf(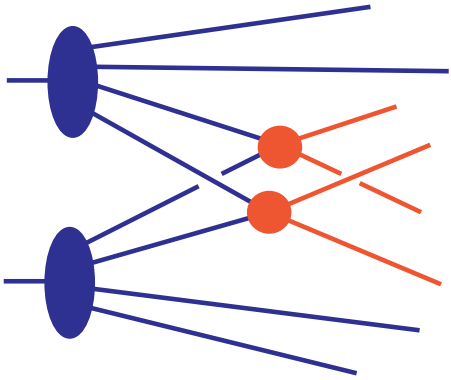 width 6 cm)}
\caption{Parton correlations can be examined experimentally by
looking for double parton scattering.}
\label{fig:doubleparton}
\end{figure}

There is some experimental information about two parton
correlations from double parton scattering. Such an
experiment is illustrated in Fig.~\ref{fig:doubleparton}. 

One could investigate this issue in a lattice calculation. One would
define a two parton distribution function $f_{(a,b)/A}(x_a,x_b,r)$
where $x_a$ and $x_b$ are the momentum fractions of two partons and
$r$ is the transverse separation between them. The simplest
uncorrelated model (ignoring even correlations due to momentum
conservation) would be that this function is the product of 
$f_{a/A}(x_a)$, $f_{b/A}(x_b)$, and a theta function that requires
that both partons are not more than a distance $R$ from each other,
$\theta(r< R)$. There must be more correlation than this. The
question is, how much?
%


\section{Partons in the pion cloud}

\begin{figure}[htb]
\vspace{9pt}
\centerline{\DESepsf(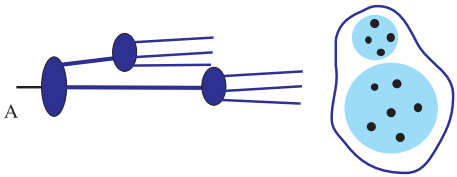 width 6 cm)}
\caption{A physical hadron may consist of bare hadrons that in turn
are made of quarks and gluons.}
\label{fig:sullivan}
\end{figure}

Figure \ref{fig:sullivan} illustrates another simple picture, usually
known as the Sullivan model \cite{sullivan}. Sometimes a proton is a
proton. Sometimes a proton is a proton plus a pion. More generally, a
proton can exist virtually as a baryon plus a meson. Each baryon or
meson consists of short distance quarks and gluons.

A formula for the parton distributions in this model is
\begin{equation}
{\blue{f_{a/A}(x;\mu)}} = 
\int_x^1 {d \xi \over \xi}\sum_{H}
{\sienna{F_{H/A}(\xi)}}\ {\green{f_{a/H}(x/\xi;\mu)}}.
\end{equation}
Here $H$ represents hadrons that might be virtual constituents of the
proton and $F_{H/A}(\xi)$ represents the distribution of these
virtual hadrons in the proton. One makes a model for
${\sienna{F_{H/A}(\xi)}}$ based on measured meson-baryon couplings.

Note that $A$ could be a proton, a ``physical proton,'' while $H$
could be a proton, a ``bare proton.'' The relation between these is,
in my opinion at least, rather murky. Nevertheless, there must be
some truth to the model. If there is some truth to it, the model
suggests that sometimes a hadron fluctuates to a much bigger size.
Again, an investigation of multiparton correlations could help to pin
the question down.


\section{Conclusion}

I conclude that lattice calculations continuing the work
\cite{latticecalc} presented at this conference could test QCD by
predicting parton distributions directly from the QCD lagrangian.
One also has the possibility of asking and answering questions about
hadron structure that go beyond what can be easily measured
experimentally.


\end{document}